# Research on Self-adaptive Online Vehicle Velocity Prediction Strategy Considering Traffic Information Fusion

Ziyan Zhang, Junhao Shen, Dongwei Yao and Feng Wu

*Abstract*—In order to increase the prediction accuracy of the online vehicle velocity prediction (VVP) strategy, a self-adaptive velocity prediction algorithm fused with traffic information was presented for the multiple scenarios. Initially, traffic scenarios were established inside the co-simulation environment. In addition, the algorithm of a general regressive neural network (GRNN) paired with datasets of the ego-vehicle, the front vehicle, and traffic lights was used in traffic scenarios, which increasingly improved the prediction accuracy. To ameliorate the robustness of the algorithm, then the strategy was optimized by particle swarm optimization (PSO) and k-fold cross-validation to find the optimal parameters of the neural network in real-time, which constructed a self-adaptive online PSO-GRNN VVP strategy with multi-information fusion to adapt with different operating situations. The self-adaptive online PSO-GRNN VVP strategy was then deployed to a variety of simulated scenarios to test its efficacy under various operating situations. Finally, the simulation results reveal that in urban and highway scenarios, the prediction accuracy is separately increased by 27.8% and 54.5% when compared to the traditional GRNN VVP strategy with fixed parameters utilizing only the historical ego-vehicle velocity dataset.

*Index Terms*—Predictive Energy Management Strategy, General Regression Neural Network, Traffic Information Fusion, Velocity Prediction

## I. INTRODUCTION

DUE to the development of hybrid technology, stricter standards are being imposed on the energy management of hybrid vehicles. The impact of the current state-of-the-art model predictive control-based energy management system is highly dependent on the precision of the anticipated future vehicle speed to calculate the future demand power every second, making it imperative to improve the prediction accuracy. The number of sensors and roadside communication devices installed on urban roads continues to rise with the implementation of vehicle-road coordination infrastructure, and the volume of real-time and historical traffic data is fast expanding. Ego-vehicles may learn about vehicles in front of them and traffic light signals through vehicle-to-vehicle and vehicle-to-infrastructure communication, which provides the data basis for predicting vehicle speed [1]. It is feasible to anticipate future vehicle velocity more precisely by analyzing data from other vehicles and roadside communication devices and combining it with the historical ego-vehicle data. This allows for the provision of consistent target demand power for energy management, which eventually results in optimal local energy usage.

Forecasting approaches based on exponential changes and data-driven methods predominate among the many methodologies for estimating vehicle velocity. Different approaches have unique application circumstances. Methods of forecasting based on exponential changes depict velocities within the forecast horizon as exponential changes. ZHANG et al. [2] proposed a vehicle speed prediction method based on exponential change, assuming that the demand torque decays exponentially to predict the vehicle velocity, and apply model predictive control (MPC) to realize power distribution. The data-driven prediction is to build a prediction model by collecting several operating condition data, and combine the current and historical working condition data to predict the vehicle velocity in the future. There are primarily two ways, Markov chain and neural network. LIU et al. [3] established a multi-step Markov chain model based on typical cycle conditions to predict vehicle velocity in a limited time domain. XIE et al. [4] used the Markov chain Monte Carlo method to predict the vehicle velocity sequence. The neural network has greater performance than the exponential change prediction approach and the Markov chain prediction approach. In addition, the neural network used in the energy management area is mainly a simple structured network with a single hidden layer, because of the requirements for real-time applications and low computing resource consumption in a vehicle, and the characteristic of a small sample size. Sun et al. [5] proposed a short-term vehicle speed predictor based on a Radial Basis Function neural network (RBFNN). With the predictor, the future demand power was calculated. Sun et al. [6] examined the prediction accuracy and computing efficiency of neural network prediction, exponential change forecast, and Markov chain prediction approaches. The findings indicate that the neural network-based speed prediction approach is more accurate. Xiang et al. [7] proposed a vehicle speed predictor based on RBFNN for predicting short-term vehicle speed, which has the advantages of fast convergence speed and low

The authors are with the Power Machinery and Vehicular Engineering Institute, Zhejiang University, Zhejiang 310027, China (e-mail: zhangziyan@zju.edu.cn; 11927048@zju.edu.cn; dwyao@zju.edu.cn; wfice@zju.edu.cn).

computational complexity. Hou et al. [8] used the RBFNN as the basis to build an adaptive prediction structure with parameters updated online and achieved a more accurate prediction effect compared with the radial basis function neural network. Some researchers used the GRNN instead of RBFNN because GRNN has a faster calculation speed and lower complexity, and it has a better performance using small samples.

Based on previous research, more and more scholars have begun to try to use traffic information to assist vehicle speed prediction. Zhang et al. [9] combined 3D map data and GPS information to use rule algorithms to predict future vehicle velocity. Gong et al. [10] used traffic information to describe the transition relation between states with mathematical models (Gipps car-following model and GAS-KINETIC model), and based on this mathematical relation, the vehicle speed sequence was predicted. He et al. [11] predicted the vehicle speed based on PSO-ELM and used the vehicle-to-vehicle communication and the vehicle-to-infrastructure communication information to calculate the safe vehicle speed as a restriction on the predicted vehicle speed to realize the correction of it. Similarly, Zhang et al. [12] used fast Fourier transform to sum the predicted vehicle speed and macroscopic velocity in the frequency domain, and used inverse fast Fourier transform to get the final result. Particularly, the algorithms proposed by the researchers weren't verified by experiments and were basically explored under simulation or dataset.

In conclusion, the majority of earlier studies relied only on historical velocity as input data, and many of them have been shown to function well under most stable operating situations. While forecast results decrease when the present speed of the ego-vehicle changes abruptly and no longer follows the same pattern as previously. It occurs, for instance, when the ego-vehicle decelerates as it approaches the traffic congestion and the traffic junction due to the front traffic light. Using traffic data and mathematical models explaining the transition relationship between states, it is possible to anticipate the sequence of vehicle speeds based on the transition relationship between states. Using the mathematical model's velocity predictions as constraints on the data-driven method's velocity predictions improves the accuracy of the current prediction. Therefore, the majority of extant research used traffic data in the form of a traffic mathematical model. In this manner, traffic lights do not assist the prediction process, which is the subject of this research.

This paper conducts a study of the ego-vehicle working conditions and the traffic information in multiple scenarios, and combines the self-adaptive optimization algorithm and the prediction algorithm to achieve an accurate predicting result online. Using the information of the previous vehicle and the traffic light, this study employs several types of fusion methods to successfully extract and combine traffic data. In addition, both the collected traffic data and the historical vehicle data are utilized as inputs to the algorithm to complete the forecast process. What's more, according to the change in the operating situations of the ego-vehicle and its environment, the optimal hyperparameters are changed automatically to better fit the driving conditions in a targeted manner.

## II. Online VVP Strategy Based On PSO-GRNN

### A. Structure of PSO-GRNN

PSO-GRNN is an algorism that combines optimization and prediction parts. PSO [13] is a population-based stochastic optimization technique, inspired by the social behavior of bird flocking or fish schooling. With PSO, the optimal parameters of GRNN can be found and updated in real-time. In this way, online prediction progress can be achieved.

GRNN [14] is a kind of forward neural network that does not use backpropagation to determine parameters. It is a form of a Radial Basis Function (RBF) network that has a fast-single learning procedure. RBF neurons compose the hidden layer at the core of the GRNN. The normal transfer function for RBF neurons is the Gaussian Function. The Gaussian Function has just one smoothing parameter σ, which defines the significance of training samples in the output results.

$$G(X, X_i) = exp(-\|X - X_i\|^2 / 2\sigma^2). \quad (1)$$

A training set includes pattern vectors $X = \begin{bmatrix} x_1 \ x_2 \ldots x_l \end{bmatrix}$ and target vectors $Y = \begin{bmatrix} y_1 \ y_2 \ldots y_p \end{bmatrix}$, which are the same as input vector and output vector. As shown in Fig. 1, the Gaussian Function processes the input vector $X$ and other pattern vectors $X_i$. $G$ represents the calculated result of the function. The function then returns the weights $w$ that indicate the proximity of the input vector to the pattern vectors.

$$w_i = exp(-\|X - X_i\|^2 / 2\sigma^2) / \sum_{j=1}^{n} exp(-\|X - X_j\|^2 / 2\sigma^2). \quad (2)$$

In addition, the weights that the hidden layer outputs amount to the final results with target vectors multiplied and indicate the relative significance of various training samples. $y_{ij}$ is the number $j$ element in $Y_i$ among the target vectors, and $y_j$ is also the number $j$ element in the output vector.

$$y_j = \sum_{i=1}^{n} w_i y_{ij} \quad (3)$$

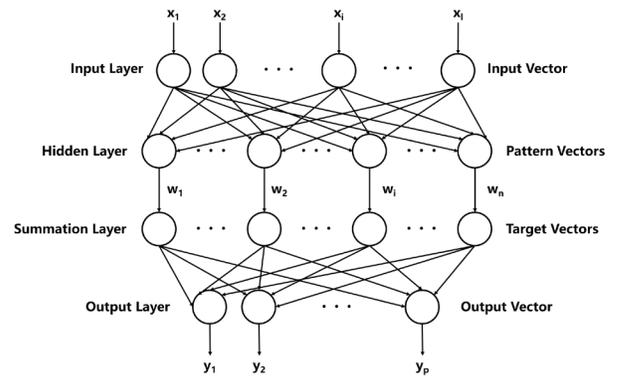

**Fig. 1.** The structure of GRNN.

Based on the concrete prediction thesis of GRNN, PSO is just used to find the optimal order (input vector number) and σ of GRNN in the strategy. The searching process consists of initializing parameters, using the specific parameters training GRNN, using cross-validation to evaluate the network performance, and comparing the performance with the target. The strategy repeats this process in a loop until the performance satisfies the requirement or the times of cycles have reached a max value. As a result, the optimal parameters can be calculated and used to predict the velocity online with PSO-GRNN.

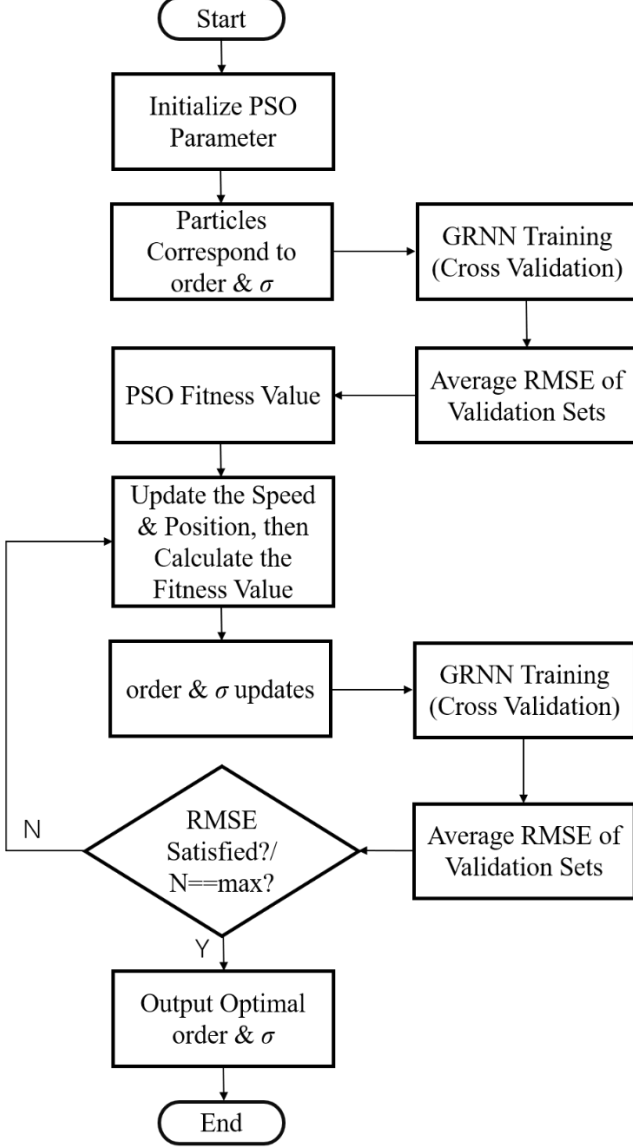

**Fig. 2.** The strategy of PSO-GRNN.

### B. Input Signals Combined with Traffic Information

In earlier velocity prediction using GRNN, the sole signal utilized as input is the historical vehicle velocity (HVV). In order to adapt to the complex driving situation and traffic conditions, various signals are considered to be fused as inputs of GRNN, including the distance between the ego-vehicle and its front vehicle (DIS), the velocity of the front vehicle (VFV), and the distance between the ego-vehicle and the traffic light (TLS).

In light of the aforementioned considerations, this study fuses the signals in three different ways: HVV+DIS, HVV+DIS+VFV, and HVV+DIS+VFV+TLS. With various signal combinations, the important task is to process the signal data in order to construct a GRNN. The remainder of this section is structured as follows: development of the GRNN structure, normalization of the data set, and the developed model based on traffic light signals.

In order to implement the online prediction method, the training dataset is updated with newly collected data every second. The first training set is obtained after $l + p$ seconds, as shown in columns in (4) and (5), and the number of the pattern vectors and their associated target vectors $M$ is 1. Following that, as time passes, the updated data is accumulated from left to right. The number of vectors in the training set at $t$ seconds is $M = t - p - l + 1$. And, as shown in (4) and (5), the pattern vectors $\begin{bmatrix} X_1 \ X_2 \cdots X_h \end{bmatrix}^T$ are in a $h \times M$ matrix, while the target vectors $\begin{bmatrix} Y_1 \ Y_2 \cdots Y_p \end{bmatrix}^T$ are in a $p \times M$ matrix. In (4), $V$ represents normalized velocity and $x$ represents one or several among normalized DIS, VFV, and TLS. $V_1$ to $V_l$ indicates the history velocity from 1 to $l$ seconds and $x_l$ represents the normalized signals at the $l$ seconds. The predicted velocity in the next $p$ seconds is given by (5) as $V_{l+1}$ to $V_{l+p}$. As a result, the training set is created. When $M$ reaches 800, the training set is reorganized with the previous data removed and the subsequent data added.

$$
\begin{bmatrix} \boldsymbol{X}_1 \\ \boldsymbol{X}_2 \\ \vdots \\ \boldsymbol{X}_{h-1} \\ \boldsymbol{X}_h \end{bmatrix} = \begin{bmatrix} V_1 & V_2 & \cdots & V_{t-p-l+1} \\ V_2 & V_3 & \cdots & V_{t-p-l+2} \\ \vdots & \vdots & \cdots & \vdots \\ V_l & V_{l+1} & \cdots & V_{t-p} \\ \boldsymbol{x_l} & \boldsymbol{x_{l+1}} & \cdots & \boldsymbol{x_{t-p}} \end{bmatrix} \quad (4)
$$

$$
\begin{bmatrix} \boldsymbol{Y}_1 \\ \boldsymbol{Y}_2 \\ \vdots \\ \boldsymbol{Y}_p \end{bmatrix} = \begin{bmatrix} V_{l+1} & V_{l+2} & \cdots & V_{t-p+1} \\ V_{l+2} & V_{l+3} & \cdots & V_{t-p+2} \\ \vdots & \vdots & \cdots & \vdots \\ V_{l+p} & V_{l+p+1} & \cdots & V_t \end{bmatrix} \quad (5)
$$

The GRNN input vector is an $h \times 1$ matrix, having the form of $[V_{(t-l+1)} \ V_{(t-l+2)} \cdots V_t \ x_t]^T$. $V$ in the vector represents the historical velocity normalized to the range from 0 to 1. When it comes to $x_t$, it might be the normalized signals such as $d$, $v_f$ and $tls$. $d$, $v_f$ and $tls$ are standardized from $D$, $V_f$ and $T$, respectively. The real distance between the ego-vehicle and its front vehicle is represented by $D$, and the real velocity of the front vehicle is represented by $V_f$, and the real distance between the ego-vehicle and the traffic light is represented by $T$.

When traffic signals are fused as HVV+DIS, the input vector

$[V_{(t-l+1)} \ V_{(t-l+2)} \cdots V_t \ d_t]^T$ is a $h \times 1$ matrix. As demonstrated in (6), $d_t$ is the normalized distance at the $t$ second and is normalized from $D_t$. Furthermore, $D$ is generated as $[D_l \ D_{(l+1)} \cdots D_t]^T$, which contains distance data from $l$ second to $t$ seconds.

$$d_t = (D_t - \min(D)) / (\max(D) - \min(D)) \quad (6)$$

When the traffic signals are combined as HVV+DIS+VFV, the input vector $[V_{(t-l+1)} \ V_{(t-l+2)} \cdots V_t \ d_t \ v_{(f_t)}]^T$ is an $(h+1) \times 1$ matrix. As indicated in (7), $v_{(f_t)}$ is normalized from $V_{(f_t)}$, and $V_f$ is generated as $[V_{(f_t)} \ V_{(f_{(t+1)})} \cdots V_{(f_t)}]^T$.

$$v_{(f_t)} = (V_{(f_t)} - \min(V_f)) / (\max(V_f) - \min(V_f)) \quad (7)$$

TLS is normalized as a numerical value compare to DIS in the signal combination HVV+DIS+VFV+TLS. According to most of the drivers' driving patterns, the driver tends to decelerate while the vehicle is approaching an intersection, hence the value of TLS is crucial to the future velocity variance. As indicated in (8), the input vector $[V_{(t-l+1)} \ V_{(t-l+2)} \cdots V_t \ d_t \ v_{(f_t)} \ tls_t]^T$ is an $(h+2) \times 1$ matrix, $tls_t$ is normalized from $T_t$, and $T$ is generated as $[T_l \ T_{(l+1)} \cdots T_t]^T$.

$$tls_t = (T_t - \min(T)) / (\max(T) - \min(T)) \quad (8)$$

### C. Parameters and Evaluation Index

In the structure itself of GRNN, its parameters are crucial in determining the neural network's prediction ability. The number of velocity variables in the vector is shown in $l$ with the previously described input vector, which is defined as order in this work. Similarly, the prediction horizon is represented by the number of variables in the output vector, donated by p. According to the reference [15], an input vector number ranging from 1 to 13 has the potential to provide a good prediction result. Furthermore, the forecast horizon is thought to be set as 5 [16]. The single parameter in the Gaussian Function is σ. According to the prior research [17], the approximate range of σ is specified as 0 to 1. Consequently, the variable parameters of the VVP models in this paper, listed as order and σ, are set as the values in the appropriate range.

The main parameters of PSO are the number of dimensions, lower bound and higher bound. The number of dimensions is defined as 2 because of the two parameters of GRNN. Similarly, the lower bound and the higher bound are also generated according to the parameters range of GRNN.

A variety of evaluation measures, such as MSE and RMSE, are used to assess the performance of VVP models. The root mean squared error (RMSE) is the most often utilized metric for evaluating VVP algorithms [18]. The RMSE denotes the variances between the predicted and actual velocity at each step, as defined in (9). Further-more, the average root mean squared error (ARMSE) is the average of the RMSEs over time, reflecting the overall performance of a VVP model.

Also, with RMSE in consideration, the neural network's performance itself can be evaluated by k-fold cross-validation. It divides the training set into $k$ parts randomly and equally, and takes one of them as the verification set to evaluate the model, and the other *(k-1)* parts as the training set to train the model. Repeat this step $k$ times, and take a different subset as the verification set each time. Finally, $k$ different models and $k$ RMSEs are obtained. The performance $E$ (average RMSEs) of these $k$ models is integrated to evaluate the advantages and disadvantages of the model in the current problem.

$$RMSE = \sqrt{1 / p \ \sum_{k=1}^{p} (V_p^k - V_d^k)^2} \quad (9)$$

$$ARMSE = 1 / M \ \sum_{step=1}^{M} RMSE \quad (10)$$

$$E = 1 / k \ \sum_{i=1}^{k} RMSE_i \quad (11)$$

## III. Simulation Analysis And Strategy Optimization

### A. Structure of Simulation Platform

SUMO [19] is a free and open-source macroscopic traffic simulator. This research uses SUMO to simulate numerous reduced urban settings. TCP/IP communication between MATLAB and SUMO was done using the connection tool traci4matlab [20]. As a consequence, the simulation is executed in SUMO, and the results are transmitted to MATLAB at each time step. In MATLAB, a GRNN-related algorithm is calculated using SUMO traffic data.

The scenarios in SUMO are constructed based on actual traffic data characteristics, with one scenario's characteristic attributes presented in Table I.

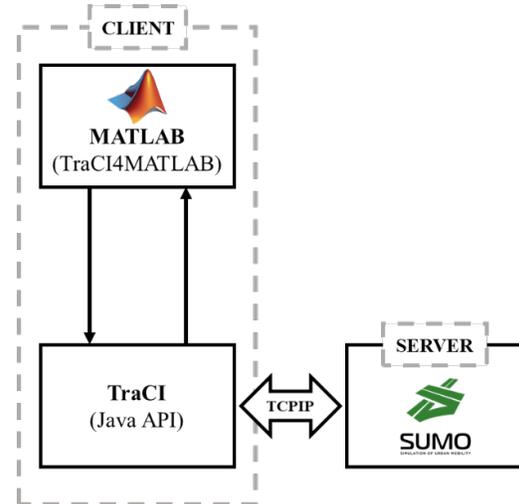

**Fig. 3.** The brief structure of the platform.

TABLE I
PARAMETERS OF DRIVING DATA

| Total mileage | Max speed | Average speed | Max positive acceleration | Max negative acceleration | Acceleration ratio | Deceleration ratio |
|---|---|---|---|---|---|---|
| 9.64(km) | 36.55(m/s) | 31.70 (m/s) | 2.60(m/s2) | -0.50(m/s2) | 46.56% | 39.34% |

*B. Comparison of Online GRNN Strategies with Different Traffic Input Signals*

In a simulated scenario, this paper compares the difference among the online VVP models with HVV, HVV+DIS, HVV+DIS+VFV, and HVV+DIS+VFV+TLS as input signals. In the VVP models, parameters are traversed with or-der from 1 to 13 and σ from 0 to 1 with a single step of 0.05. With more information involved in the model, the optimal prediction performance of HVV+DIS+VFV+TLS is shown in Table 2. Compared with HVV, HVV+DIS, and HVV+DIS+VFV, the ARMSE goes down to 1.149. As a result, the performance improves by 14.9%.

TABLE II
OPTIMAL ARMSES UNDER VARIOUS SIGNAL COMBINATION INPUTS IN URBAN SCENARIOS

| Inputs | HVV | HVV + DIS | HVV + DIS + VFV | HVV + DIS + VFV+TLS |
|---|---|---|---|---|
| ARMSE | 1.351 | 1.282 | 1.173 | 1.149 |
| Improvement | - | 5.1% | 13.2% | 14.9% |

With four kinds of traffic information fused as GRNN inputs, the more accurate predicted results are obtained and depicted with driving data of the target vehicle in every time step. The general vehicle velocity data is characterized by many variations in speed. Thus, it's more difficult to predict the uncertain velocity in the future, compared with stable velocity like in a highway scenario. With more signals especially traffic light signals fused, the optimal prediction performance of GRNN with HVV+DIS+VFV+TLS is shown in Fig. 4, compared with the optimal performance of GRNN with HVV in Fig. 3.

As can be seen, the prediction performance is better in fluctuation situations with multi-signals fused, especially when the actual velocity appears to be different from the training samples that exist in the dataset. For instance, when time flows from 440s to 540s, the prediction results of GRNN with HVV+DIS+VFV+TLS have a smaller velocity deviation than the actual velocity compared to the results in Fig. 4. As shown in Fig. 5, RMSE of the VVP strategy with fused signals is smaller than others when actual velocity fluctuates.

With more traffic signals fused into the VVP strategy, prediction performance is improved. While in the GRNN with HVV+DIS+VFV+TLS, its internal parameters are the keys to improving the prediction effect. The traversal results of different combinations of order and σ are shown in Table III and IV, and they have a great impact during the overall operating process. As a result, the ARMSE of GRNN with the optimal σ being 0.1 and order being 2 is the meanest in all optimal ARMSEs under different combinations.

With the analysis of simulation results, the fused traffic signals play a great role in the improvement of prediction results. And also, the optimal parameters of GRNN with HVV+DIS+VFV+TLS inputs can be traversed and evaluated at the end of the operating process, but they are fixed during the whole process, and they can't be changed dynamically in real-time to adapt to the changeable operating situations. Therefore, PSO, the online parameter optimizing algorithm, is used in the subsequent VVP strategy.

*C. Self-adaptive Optimization of the Online GRNN Strategy using PSO*

Using PSO, the order and the σ in each step can be calculated constantly and eventually converge to rather accurate values. The evaluation index of the parameters' accuracy is the k-fold cross-validation result *E*. With cross-validation of the GRNN with specific optimized parameters by PSO, the neural network's performance can be achieved. Consequently, the optimal parameters are calculated with a rather small value of $E$. However, this kind of optimization method only gives a pair of relatively accurate values rather than the optimal values because of the limit of calculation time and the uncertainty of future velocity.

TABLE III
OPTIMAL ARMSES UNDER DIFFERENT ORDERS WITH HVV+DIS+VFV+TLS INPUTS OVERALL

| Order | 1 | 2 | 3 | 4 | 5 | 6 | 7 | 8 | 9 | 10 | 11 | 12 | 13 |
|---|---|---|---|---|---|---|---|---|---|---|---|---|---|
| ARMSE | 1.176 | 1.149 | 1.153 | 1.177 | 1.187 | 1.194 | 1.211 | 1.210 | 1.210 | 1.212 | 1.220 | 1.236 | 1.255 |

TABLE IV
OPTIMAL ARMSES UNDER DIFFERENT σ WITH HVV+DIS+VFV+TLS INPUTS OVERALL

| σ | 0.05 | 0.10 | 0.15 | 0.20 | 0.25 | 0.30 | … | 0.75 | 0.80 | 0.85 | 0.90 | 0.95 | 1.00 |
|---|---|---|---|---|---|---|---|---|---|---|---|---|---|
| ARMSE | 1.176 | 1.149 | 1.175 | 1.219 | 1.250 | 1.286 | … | 1.652 | 1.700 | 1.746 | 1.788 | 1.823 | 1.853 |

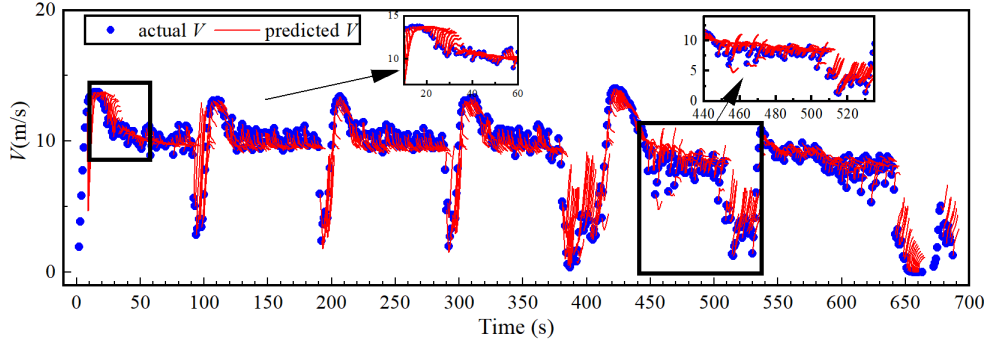

**Fig. 4.** The single-step prediction velocity distribution of GRNN with HVV in urban scenarios.

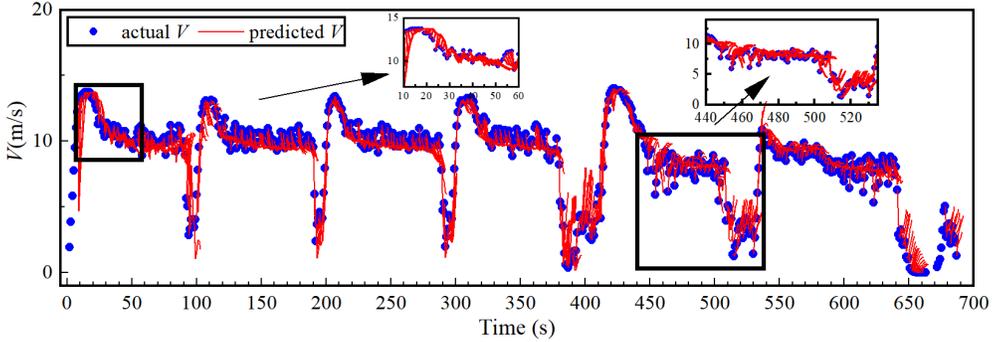

**Fig. 5.** The single-step prediction velocity distribution of GRNN with HVV+DIS+VFV+TLS in urban scenarios.

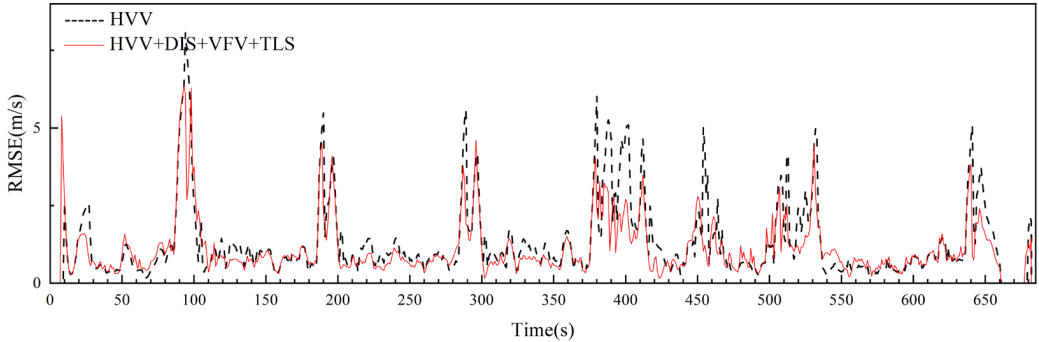

**Fig. 6.** Single-step RMSE distributions corresponding to the optimal ARMSE in urban scenarios.

Taking an urban scenario as an example, with an optimized GRNN strategy, which uses HVV+DIS+VFV+TLS as input signals, this paper optimizes the strategy with PSO continuously, in order to achieve the online self-adaptation. However, as a comparison, the GRNN strategy with HVV+DIS+VFV+TLS inputs is with random fixed parameters, which aren't their optimal ones overall. For example, order can be set as 7, and σ can be set as 0.50, which are in the medium of the range of the parameters. The ARMSEs of these two strategies are shown in Table V. The ARMSE of the online self-adaptive PSO-GRNN strategy has improved by 13.3%. What's more, it has exceeded 84.2% ARMSEs of the GRNN strategy with fixed parameters, whose ranges are 1-13 and 0-1. As shown in Fig. 6, the RMSE of the PSO-GRNN strategy is smaller than the GRNN strategy with fixed parameters when the actual velocity fluctuates. As a result, it proves the effectiveness of the PSO-GRNN strategy.

TABLE V
ARMSEs UNDER PSO-GRNN AND GRNN WITH FIXED
PARAMETERS

| Inputs | *HVV + DIS + VFV+TLS* | *HVV + DIS + VFV+TLS* |
|---|---|---|
| Parameters | Fixed (7/0.50) | PSO |
| ARMSE | 1.448 | 1.255 |
| Improvement | - | 13.3% |

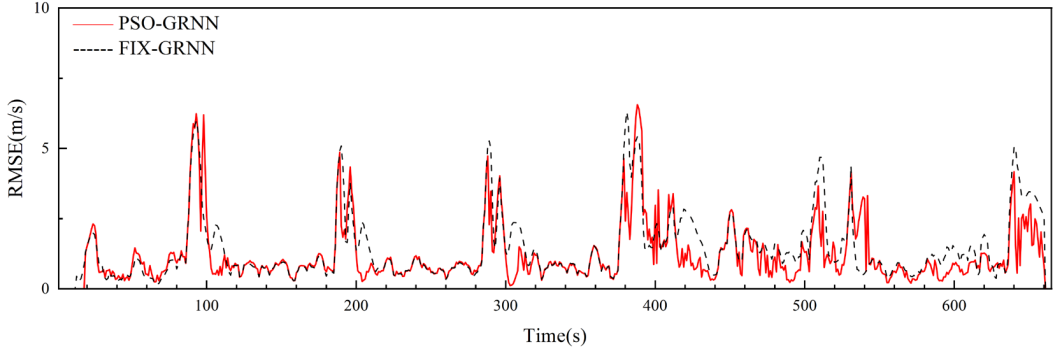

**Fig. 7.** Single-step RMSE distributions corresponding to the optimal ARMSE in PSO-GRNN and GRNN with fixed parameters.

## IV. Verification of The Optimal VVP Strategy

### A. Verification with Multiple Scenarios

With newly simulated scenarios based on real traffic data, the former algorithm can be verified. Several urban and highway scenarios are built. In urban scenarios, the self-adaptive online VVP strategy with HVV+DIS+VFV+TLS inputs is verified using different urban scenarios, including high traffic density scenarios and low traffic density scenarios. Also in highway scenarios, the self-adaptive online VVP strategy with HVV+DIS+VFV inputs is verified with high-speed scenarios and middle-speed scenarios.

### B. Verification of Self-adaptive Online VVP Strategy in Urban Scenarios

Based on the built test scenarios, this paper verifies the effectiveness of the self-adaptive online PSO-GRNN strategy with HVV+DIS+VFV+TLS inputs. The traditional fixed online GRNN strategy with HVV inputs is used as a benchmark in the same scenarios. As a result, the summarized ARMSE results of the comparison of the strategies based on different test scenarios are shown in Table VI.

TABLE VI
ARMSEs under PSO-GRNN and GRNN with Fixed Parameters in Urban Scenarios

| Inputs | HVV (test1) | HVV + DIS + VFV+TLS (test1) | HVV (test2) | HVV + DIS + VFV+TLS (test2) |
|---|---|---|---|---|
| Parameters | Fixed (7/0.50) | PSO | Fixed (7/0.50) | PSO |
| ARMSE | 1.451 | 1.062 | 1.566 | 1.117 |
| Improvement | - | 26.8% | - | 28.7% |

It can be seen that the performance of the self-adaptive online PSO-GRNN strategy with HVV+DIS+VFV+TLS inputs has all improved a lot in different operating situations, compared with the fixed online GRNN strategy with HVV inputs, whose parameters are in the middle of the traversal range. Then the overall ARMSEs of the proposed strategy in two tests separately overperform 26.8% and 28.7%. Also, the results of the proposed strategy have exceeded 100% of the traversal results of the traditional strategy both in test1 and test2, which can demonstrate the efficiency of the self-adaptive online PSO-GRNN strategy in urban scenarios.

### C. Verification of Self-adaptive Online VVP Strategy in Highway Scenarios

Same as in urban scenarios, the effectiveness of the self-adaptive online PSO-GRNN strategy with HVV+DIS+VFV inputs is verified in highway scenarios. The benchmark used as a comparison is also the fixed online GRNN strategy with HVV inputs. Thus, the proposed strategy is tested in different highway scenarios and its effectiveness is verified in Table VII.

TABLE VII
ARMSEs under PSO-GRNN and GRNN with Fixed Parameters in Highway Scenarios

| Inputs | HVV (test3) | HVV + DIS + VFV (test3) | HVV (test4) | HVV + DIS + VFV (test4) |
|---|---|---|---|---|
| Parameters | Fixed (7/0.50) | PSO | Fixed (7/0.50) | PSO |
| ARMSE | 1.243 | 0.580 | 2.268 | 1.005 |
| Improvement | - | 53.3% | - | 55.7% |

Also, the results reveal that the self-adaptive online PSO-GRNN strategy with HVV+DIS+VFV inputs has improved the prediction accuracy by 53.3% and 55.7%, compared with the fixed online GRNN strategy with HVV inputs, whose order is 7 and σ is 0.50. In the prediction results of various traversal combinations of the parameters in the fixed online GRNN strategy, 100% and 99.2% of them have been exceeded by the results of the self-adaptive online PSO-GRNN strategy, corresponding to test3 and test4.

It can be seen that the prediction improvement ratios of urban scenarios and highway scenarios are rather different, which is the result of more regular operating situations with fewer fluctuations in velocity in highway scenarios. Thus, the proposed strategy can make the most of regularity in a scenario, the more regular the scenario is, the more effective the strategy is.

## V. CONCLUSION

The proposed self-adaptive online PSO-GRNN VVP strategy considering traffic information uses front vehicle signal and front traffic light signal to achieve the target: the improvement of the prediction process itself (GRNN) with traffic information. First, this paper proposes an online fusing traffic signal strategy to improve prediction accuracy. Second, PSO has been used to achieve the goal of self-adaptive prediction with the signal-fused strategy. In addition, the verification of the proposed strategy demonstrated its effectiveness in most scenarios in signal fusing and self-adaptive online optimization. In the final test results, the average ARMSE of the two tests is separately improved by 27.8% in urban scenarios and by 54.5% in highway scenarios, which means more accurate speed prediction results provide more optimization space for subsequent energy management. What's more, fusing more traffic signals into the GRNN to improve prediction accuracy, such as traffic light state signals, is considered to be our future work.

## APPENDIX

| | |
|---|---|
| VVP | Vehicle Velocity Prediction |
| GRNN | General Regressive Neural Network |
| PSO | Particle Swarm Optimization |
| PSO GRNN | Particle Swarm Optimization-General Regressive Neural Network |
| MPC | Model Predictive Control |
| RBFNN | Radial Basis Function Neural Network |
| GPS | Global Position System |
| PSO-ELM | Particle Swarm Optimization-Extreme Learning Machine |
| RBF | Radial Basis Function |
| HVV | Historical Vehicle Velocity |
| DIS | Distance between the Ego-vehicle and Its Front Vehicle |
| VFV | Velocity of the Front Vehicle |
| TLS | Distance between the Ego-vehicle and the Traffic Light |
| MSE | Mean Squared Error |
| RMSE | Root Mean Squared Error |
| ARMSE | Average Root Mean Squared Error |
| order | Number of Velocity Elements in the Input Vector |
| $\sigma$ | Hyper-parameter of the Gaussian Function |
| $X$ | Input Vector |
| $Y$ | Output Vector |
| $G$ | Results of the Gaussian Function |
| $w$ | Weight of the Input Vector |
| $l$ | Number of History Velocities in the Input Vector |
| $p$ | Number of Future Velocities in the Output Vector |
| $M$ | Number of Vectors in the Training Set |
| $h$ | Number of Elements in the Input Vector |
| $D$ | Real Distance between the Ego-vehicle and Its Front Vehicle |
| $d$ | Normalized $D$ |
| $V_f$ | Real Velocity of the Front Vehicle |
| $v_f$ | Normalized $V_f$ |
| $T$ | Real Distance between the Ego-vehicle and the Traffic Light |
| $tls$ | Normalized $T$ |
| $E$ | Average RMSEs of k Models in K-fold Cross-validation |

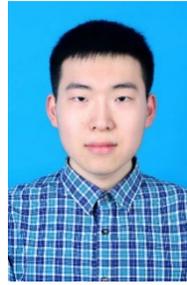

**Ziyan Zhang** received the B.Eng. degree in vehicle engineering from the College of Automotive Engineering, Chongqing University, Chongqing, China, in 2020. He is currently pursuing the M.Eng. degree in energy engineering with the School of Energy Engineering, Zhejiang University, Zhejiang, China.

His research interests include energy management strategies for hybrid electric vehicles, and the simulation of intelligent transportation systems with connected vehicles.

**Junhao Shen**, photograph and biography not available at the time of publication.

**Dongwei Yao**, photograph and biography not available at the time of publication.

**Feng Wu**, photograph and biography not available at the time of publication.